# Multivariate Dual to Ratio Type Estimators using Arithmetic, Geometric and Harmonic Means in Simple Random Sampling

Rajesh Singh, Prayas Sharma
Department of Statistics, Banaras Hindu University
Varanasi-221005, India

**Abstract**

Auxiliary variable is extensively used in survey sampling to improve the precision of estimates. Whenever there is availability of auxiliary information, we want to utilize it in the method of estimation to obtain the most efficient estimator. In this paper using multi-auxiliary information we have proposed estimators based on arithmetic, geometric and harmonic mean. It was also shown that estimator based on harmonic and geometric means are more biased than estimator based on arithmetic mean under certain conditions. However, the MSE of all three estimators are same up to the first order of approximation.

## 1. Introduction.

Multivariate has been widely discussed and applicable in finite population sampling literature. Olkin (1958) has considered the use of multi-auxiliary variables when they are positively correlated with the variable under study to build-up a multivariate ratio estimator of population mean $\overline{Y}$. Srivenkataramana (1980) first time proposed the dual to ratio estimator for estimating the population mean. Singh and Tailor (2005), Sharma and Tailor (2010) proposed some ratio cum-dual to ratio estimators for the estimation of finite population mean of study variable y. In addition, Javid Shabbir (2006) suggested a dual to variance ratio type estimator. Let U =($U_1$,$U_2$,.....$U_N$ ) be the finite population of size N out of which a sample of size n is drawn using simple random sampling without replacement technique. Let y and x be the study and the auxiliary variables respectively and y is positively correlated with x. Let $\overline{Y} = \frac{1}{N}\sum_{i=1}^{N} Y_i$ and $\overline{X} = \frac{1}{N}\sum_{i=1}^{N} X_i$ be the population mean of study and auxiliary variables and $\overline{y} = \frac{1}{N}\sum_{i=1}^{n} y_i$ and $\overline{x} = \frac{1}{N}\sum_{i=1}^{n} x_i$ be the respective sample means.

Motivated by Olkin (1958) we propose a dual to ratio type estimator based on weighted arithmetic mean of $r_i \overline{X}_i$'s and is given as

$$\overline{y}_{ap} = \sum_{i=1}^{k} \alpha_i r_i \overline{X}_i \tag{1.1}$$

Where **(i)** $\alpha_i$'s are weights such that $\sum_{i=1}^{k} \alpha_i = 1$, **(ii)** $\overline{X}_i$'s are the population means of the auxiliary variables and assumed to be known and **(iii)** $r_i = \frac{\overline{y}}{\overline{x}_i^*}$,

Where $\overline{x}_i^* = \frac{N\overline{X} - nx_i}{N - n}$ or $\overline{x}_i^* = (1+g)\overline{X} - gx_i$ i=1,2,3….N,

Which usually gives $\overline{x}_i^* = (1+g)\overline{X} - g\overline{x}$, where $g = \frac{n}{N-n}$

Following Olkin's estimator, in recent years, several other estimators have been proposed using multi- auxiliary variables. Singh (1967) has extended Olkin's estimator to the case where auxiliary variables are negatively correlated with the variable under study. Srivastava (1965), and Rao and Mudholkar (1967) have suggested estimators, where some of the variables are positively and others are negatively correlated with the variable under study. Here, the main objective of presenting these estimators is to reduce the bias and mean square errors.

Motivated by Singh (1965, 1967) and Singh et al. (2007), we propose a dual to ratio estimator as

$$\overline{y}_s = \prod_{i=1}^{k} r_i \overline{X}_i \tag{1.2}$$

We also propose two alternative estimators based on geometric mean and harmonic mean, as

$$\overline{y}_{gp} = \prod_{i=1}^{k} \left(r_i \overline{X}_i\right)^{\alpha_i} \tag{1.3}$$

$$\overline{y}_{hp} = \left(\sum_{i=1}^{k} \frac{\alpha_i}{r_i \overline{X}_i}\right)^{-1} \tag{1.4}$$

Such that $\sum_{i=1}^{k}\alpha_i = 1$

These estimators are based on the assumptions that the auxiliary variables are positively correlated with Y. Let $\rho_{ij}$ (i=1,2,…k; j=1,2,…k ) be the correlation coefficient between $X_i$ and $X_j$ and $\rho_{0i}$ be the correlation coefficient between Y and $X_i$.

## 2. BIAS AND MSE OF THE ESTIMATES

To obtain the bias and MSE's of the estimators up to first order of approximation, we obtain

$\bar{y} = \bar{Y}(1+e_0) \quad \bar{x}_i = \bar{X}_i(1+e_i)$

Such that $E(e_i) = 0$

Where,

$E(e_i^2) = C_i^2$, For i=0,1,2…..

$E(e_i e_j) = C_{0i}$, For (i,j)=0,1,2…..

$V(\bar{y}) = \bar{Y}E(e_0^2)$

In the same way $C_{0i}$ and $C_{ij}$ are defined.

Further, let $\underset{\sim}{\alpha}' = (\alpha_1, \alpha_2, ..., \alpha_k)$ and $C = \lfloor C_{ij} \rfloor_{k\times k}$, (i=1,2,…k; j=1,2,…k )

Using Taylor's series expansion under the usual assumption, expanding right hand side of equation (1.1) in terms of e's we obtain,

$$\bar{y}_{ap} = \sum_{i=1}^{k}\alpha_i \bar{Y}(1+e_0)(1+e_i)^{-1}$$

$$= \bar{Y}\sum_{i=1}^{k}\alpha_i \left[(1+e_0)(1-e_i + e_i^2)\right]$$

$$= \bar{Y}\sum_{i=1}^{k}\alpha_i \left[1+e_0 - e_i + e_i^2 - e_0 e_i\right] \tag{2.1}$$

Subtracting $\bar{Y}$ from both the sides of equation (2.1) and then taking expectation of both sides, we get the bias of the estimator $\bar{y}_{ap}$ to the first order of approximation as

$$B(\bar{y}_{ap}) = \bar{Y}\left[g^2\sum_{i=1}^{k}\alpha_i C_i^2 + g\sum_{i=1}^{k}\alpha_i C_{0i}\right] \tag{2.2}$$

Subtracting $\bar{Y}$ from both the sides of equation (2.1) taking square and then taking expectation of both sides, we get the bias of the estimator $\bar{y}_{ap}$ to the first order of approximation as

$$MSE\left(\bar{y}_{ap}\right)=\bar{Y}^2\left[C_0^2+\sum_{i=1}^{k}\alpha_i^2C_i^2+2g^2\sum\sum\alpha_i\alpha_jC_{ij}+2g\sum_{i=1}^{k}\alpha_iC_0C_i\right] \tag{2.3}$$

In the same way using the Taylor series expansion under the usual assumptions, from equation (1.3) we obtain

$$\bar{y}_{gp}=\bar{Y}\prod_{i=1}^{k}\left[1+e_0-\alpha_i\left(e_i+e_0e_i\right)+\frac{\alpha_i\left(1+\alpha_i\right)}{2}\left(e_i^2+e_0e_i^2\right)\right] \tag{2.4}$$

And

$$\bar{y}_{hp}=\bar{Y}\left[1+e_0-\sum_{i=1}^{k}\alpha_ie_i-\sum_{i=1}^{k}\alpha_ie_0e_i+\left(\sum_{i=1}^{k}\alpha_ie_i\right)^2+\left(\sum_{i=1}^{k}\alpha_ie_i\right)^2e_0\right] \tag{2.5}$$

To calculate the bias and mean square error, we considered the terms having powers up to second degree only as the calculations become more complicated when the higher order terms are included.

So, from equation (2.4) and (2.5), the bias and mean square error of the estimates up to 0(1/n) are obtained as:

$$B\left(\bar{y}_{gp}\right)=\bar{Y}\left[g^2\sum_i\frac{\alpha_i\left(\alpha_i+1\right)C_i^2}{2}+g\sum\sum\alpha_i\alpha_jC_{ij}+g\sum_i\alpha_iC_{0i}\right] \tag{2.6}$$

$$MSE\left(\bar{y}_{gp}\right)=\bar{Y}^2\left[C_0^2+g^2\sum_{i=1}^{k}\alpha_i^2C_i^2+2g\sum_{i=1}^{k}\alpha_iC_{0i}+2g^2\sum\sum\alpha_i\alpha_jC_{ij}\right] \tag{2.7}$$

And

$$B\left(\bar{y}_{hp}\right)=\bar{Y}\left[\left(g\sum_i\alpha_iC_i\right)^2+g\sum_i\alpha_iC_{0i}\right] \tag{2.8}$$

$$MSE\left(\bar{y}_{hp}\right)=\bar{Y}^2\left[C_0^2+g^2\sum_{i=1}^{k}\alpha_i^2C_i^2+2g\sum_{i=1}^{k}\alpha_iC_{0i}+2g^2\sum\sum\alpha_i\alpha_jC_{ij}\right] \tag{2.9}$$

We see that MSE's of these estimators are same but the biases are different. In general

$$MSE\left(\bar{y}_{ap}\right)=MSE\left(\bar{y}_{gp}\right)=MSE\left(\bar{y}_{hp}\right) \tag{2.10}$$

We know that in case of univariate the usual ratio type $\bar{y}_R$ estimator for the $i^{th}$ auxiliary variable is superior to the mean per unit estimator $\bar{y}$, when

$$\frac{C_0}{C_i}\rho_{0i}>\frac{1}{2} \tag{2.11}$$

Comparing the variance of $\bar{y} = C_0^2 \bar{Y}^2$ with the mean square error of all the three estimators, we note that the ratio estimator given in (1.1), (1.2) and (1.3) are more efficient than $\bar{y}$.

**3. Comparison of Biases**

The biases may be either positive or negative. So, for comparison, we have compared the absolute biases of the estimates when these are more efficient than the sample mean. The bias of the estimator of arithmetic mean is smaller than that of geometric mean.

$$\left|B\left(\bar{y}_{gp}\right)\right| > \left|B\left(\bar{y}_{ap}\right)\right| \tag{3.1}$$

Simplifying (3.1), we observe that condition (3.1) holds true, if

$$\left[\left(g^2\sum_{i=1}^{k}\alpha_i^2 C_i^2 + 2g\sum\sum\alpha_i\alpha_j C_{ij}\right) - g^2\sum_{i=1}^{k}\alpha_i C_i^2\right] > 0 \tag{3.2}$$

Similarly,

$$\left|B\left(\bar{y}_{hp}\right)\right| > \left|B\left(\bar{y}_{gp}\right)\right|$$

$$\bar{Y}\left[\left(g\sum_i \alpha_i C_i\right)^2 + g\sum_i \alpha_i C_{0i}\right] > \bar{Y}\left[g^2\sum_i \frac{\alpha_i(\alpha_i+1)C_i^2}{2} + g\sum\sum\alpha_i\alpha_j C_{ij} + g\sum_i \alpha_i C_{0i}\right] \tag{3.3}$$

Simplifying equation (3.3), we observe that equality holds for

$$g > \frac{1}{2} \tag{3.4}$$

When conditions, (3.2) and (3.4) holds, we see that harmonic estimator is more bias than geometric estimator and geometric estimator is more biased than arithmetic estimator. Hence we may conclude that under the aforesaid conditions

$$\left|B\left(\bar{y}_{hp}\right)\right| > \left|B\left(\bar{y}_{gp}\right)\right| > \left|B\left(\bar{y}_{ap}\right)\right|$$

**4. Empirical Study**

We have applied the traditional and proposed estimator on the data of Apple production amount in 1999 (as interest of variate) and number of apple trees in 1999 (as first auxiliary variate), apple production amount in 1998 (as second auxiliary variate) of 204 villages in Black Sea Region of Turkey (Source:Institute of Statistics, Republic of Turkey).

**Table 4.1: Data Statistics**

| | | |
|---|---|---|
| N=204 | n=50 | g=0.3246 |
| $S_{x1}$= 45402.78 | $S_{x2}$= 2521.40 | $S_y$=2389.76 |
| $\bar{X}_1$ = 26441 | $\bar{X}_2$ = 1014 | $\bar{Y}$ =966 |
| $S_{x1y}$=77372777 | $S_{yx2}$ =5684276 | $\rho_{yx1}$ =0.71 |
| $\rho_{yx2}$ = 0.94 | $\rho_{x1x2}$ =0.83 | $B_1$=0.04 |
| $B_2$= 0.89 | | |

Biases and MSE's of different estimators under comparison, based on the above data are given in Table 4.1.

**Table 4.2: Bias and MSE of different estimators**

| Estimators | Auxiliary variables used | Absolute Bias | MSE |
|---|---|---|---|
| $\bar{y}_{st}$ | none | 0 | 5710952 |
| Ratio $\bar{y}\left(\frac{\bar{X}_1}{\bar{x}^*}\right)$ | $X_1$ | 649.0 | 4165443 |
| Ratio $\bar{y}\left(\frac{\bar{X}_2}{\bar{x}^*}\right)$ | $X_2$ | 1190 | 2802810 |
| Suggested $\bar{Y}_{ap}$ | $X_1$ and $X_2$ | 3389 | 4239.70 |
| Suggested $\bar{y}_{gp}$ | $X_1$ and $X_2$ | 3501 | 4239.70 |
| Suggested $\bar{y}_{hp}$ | $X_1$ and $X_2$ | 3690 | 4239.70 |

## 5. Conclusion

From the table 4.2, we observed that the dual to ratio estimator based on arithmetic mean is less biased. However, the mean square errors of the estimators $\bar{y}_{ap}$, $\bar{y}_{gp}$ and $\bar{y}_{hp}$ are same. Hence for this data set, we conclude that when more than one, auxiliary variables are used for estimating the population parameters, it is better to use arithmetic mean as an estimator in case of simple random sampling

## References

Olkin, I. (1958): Multivariate ratio estimation for finite population. Biometrica, 45. 154-165.

Rao, P.S.R.S. and Mudholkar, G.S. (1967): Generalized multivariate estimator for the mean of a finite population. Jour. Amer. Stat. Assoc., 62, 1009-1012.

Shabbir, J. (2006): A dual to variance ratio-type estimator in simple random sampling. Proc. Pakistan Acad. Sci. 43(4):279-283.

Sharma, B. and Tailor, R. (2010): A new ratio-cum-dual to ratio estimator of finite population Mean in simple random sampling. Global Journal of Science Frontier Research, 10, 1, 27-31.

Singh, R., Chauhan, P., Sawan, N. and Smarandache, F. (2007): Auxiliary information and a priori values in construction of improved estimators, Renaissance High press, USA, 2007.

Singh, H.P. and Tailor, R. (2005), Estimation of finite population mean using known Correlation coefficient between auxiliary characters. Statistica, Anno vol. LXV (4), pp. 407-418.

Singh, M. P. (1965): On the estimation of ratio and product of population parameter. Sankhya 27 B, 321-328.

Singh, M.P. (1967): Multivariate product method of estimation for finite populations. J. Indian Soc. Agri. Statist., 19, 1-10.

Srivastava, S.K. (1965): An estimate of the mean of a finite population using several auxiliary variables. J. Indian Soc. Agri. Statist., 3, 189-194.

Srivenkataramana, T. (1980): A dual to ratio estimator in sample surveys. Biometrika, vol. 67 (1), pp. 199 - 204.